\documentclass{elsart}

\usepackage{graphicx}
\usepackage{amssymb}

\begin{document}
\begin{frontmatter}

\title{MONTE CARLO SIMULATION AND STATISTICAL ANALYSIS OF THE EFFECT OF CODING TABLE SPECIFICITY ON GENETIC INFORMATION CODING}
\author{E. Gultepe}
\address{Physics Department, Northeastern University, Boston, Massachusetts, 02115, USA}
\author{M.~L. Kurnaz\corauthref{BU}}
\address{Department of Physics, Bogazici University, 34342
Bebek Istanbul, Turkey}
\ead{levent.kurnaz@boun.edu.tr}
\corauth[BU]{Corresponding Author}

\begin{abstract}
We present a computer simulation, which is inspired by Penna model, to help understanding the effect of genetic coding tables on population dynamics. To represent populations we used real and  artificial gene sequences in this model. We coded these sequences using different amino acid tables in Nature, the standard table as well as the tables which are used by mithocondria and some eukaryotes. Contrary to common belief we find that  the standard code table which is used in most organisms in Nature, does not give the most resilient coding against point mutations. 
\end{abstract}

\end{frontmatter}

\section{INTRODUCTION\protect\\ }
\label{sec:level1}
Modeling population dynamics has been popular in physics community for a while mostly because complexity of the system bears the necessity of the statistical and computational tools. Physicists brought the models which can give rise to computational simulations together with \textit{looking for the simplest solution} approach into this subject. Among all others \cite{Heumann1995,Mueller1996,Cui2000,Medeiros2001}, Penna model \cite{Penna} is the most extensive simulation scheme used in population dynamics. Simulations for population dynamics usually take many different factors of live into account. In real life it is very difficult to have only one of the aspects of life count where we neglect the effects of everything else. For example in real life you cannot say that the only cause for death is point mutations because then you have to keep the individuals in the system from dying of "old age" or of malnutrition or of fighting amongst the members. In simulations it is much easier to ignore all these facts of life and concentrate only on one simple aspect. In this work we have neglected all the other aspects of life and concentrated only on the effect of point mutations on survivability of the individuals.

Genetic information of all living organism (except some viruses) is stored in DNA. The segment of DNA which contains necessary information to produce a specific protein is called \textit{gene}. A real gene is composed of two different parts: a coding portion and a non-coding portion. The coding part, exon, is responsible for protein synthesis whereas the rest, intron, does not code  a protein and the purpose of this part is not clearly understood yet.

The information in DNA is coded by using four different types of monomers adenine (A), guanine (G), cytosine (C) and thymine(T). These monomers are the letters of the genetic alphabet and they construct 3-letters long words, \textit{codons}. Every codon on DNA codes an amino-acid during the protein synthesis. There are $4^3$, 64 possible combination of codons available on DNA. However; in Nature there are only 20 amino acids available for protein coding and as a result there is no one-to-one codon-amino acid correspondence. The table which determines how the codons are mapped into the amino acids is called \textit{amino acid table} or \textit{genetic coding table}.

For a long time it was believed that the amino acid table of Nature was universal, \textit{Standart Genetic Code} (SGC). However; we now know that there are few exceptions in codon usage which  are confined to mitochondria and certain protozoa \cite{Weaver2002}. In these particular tables the same number of amino acids is used, but some of these amino acids are coded by different codons. For example, in \textit{SGC} , the codon ``AUA" codes Isoleucine, the codon ``UGA" codes Stop, and the codons ``AGA" and ``AGG" code Arginine . However, in the table of \textit{Vertebrate Mitochondrial Code}, they code Methionine, Tryptophan, and Stop respectively \cite{NCBIcodes}. 

Recently, we have developed a Monte Carlo simulation model \cite{Gultepe05} inspired by the Penna model to investigate the significance of the number of amino acids in population dynamics. In that model, each individual was represented by a human cytokine gene sequence and mutation was assumed to be the only cause of death by eliminating all other effects. In that study, it has been shown that for maximum tolerance against mutations, the number of amino acids which codes the genetic information, is bounded between $20$ to $24$. The number of amino acids used in Nature, $21$ (20 amino acids plus the Stop codon), is in this optimum range.

In this paper, we used the same model to investigate the endurance of different amino acid tables against point mutations neglecting any other causes of death.

\section{COMPUTATIONAL METHOD\protect\\ }
\label{sec:level2}

In our model, an individual was represented by three different gene sequences. First, we have used a real gene from Nature \textit{human cytokine} (LD78 Homo sapiens blood lymphocyte gene on the DNA 17$^{th}$ chromosome) \cite{gene} same as in the previous model. This gene is playing an important role in immune system of human body, hence any problem in generating this gene is lethal. Afterwards we used the same model on different mammalian genes, in this paper we also report the results for human ARNT gene (Homo sapiens aryl hydrocarbon receptor nuclear translocator, transcript variant 1, mRNA) as an example.

Next, we have created an artificial human gene, the average human gene, which basically reflects the codon usage frequency found in Homo sapiens. This gene consists of 1000 nucleotides (frequency is taken as an integer value per one thousand nucleotides given in \cite{Kazusa}) in a randomly chosen sequence. This artificial gene is considered to represent the whole human genome. This assures that the results we have obtained are not specific to the human cytokine gene but for whole genome.

For the sake of simplicity, we have not included reproduction and we also have neglected all other effects causing death, except mutation. Using this model, we have investigated the effects of mutations on the population size. A mutation in this model was taken as a change of one nucleotide in the gene.  It can be either lethal or harmless depending on whether it causes a change in the amino acid chain or not. We kept all mutation rates equal like in the Jukes-Cantor mutation scheme \cite{Jukes21920}. 

\begin{figure}[!]

\includegraphics[width=14cm]{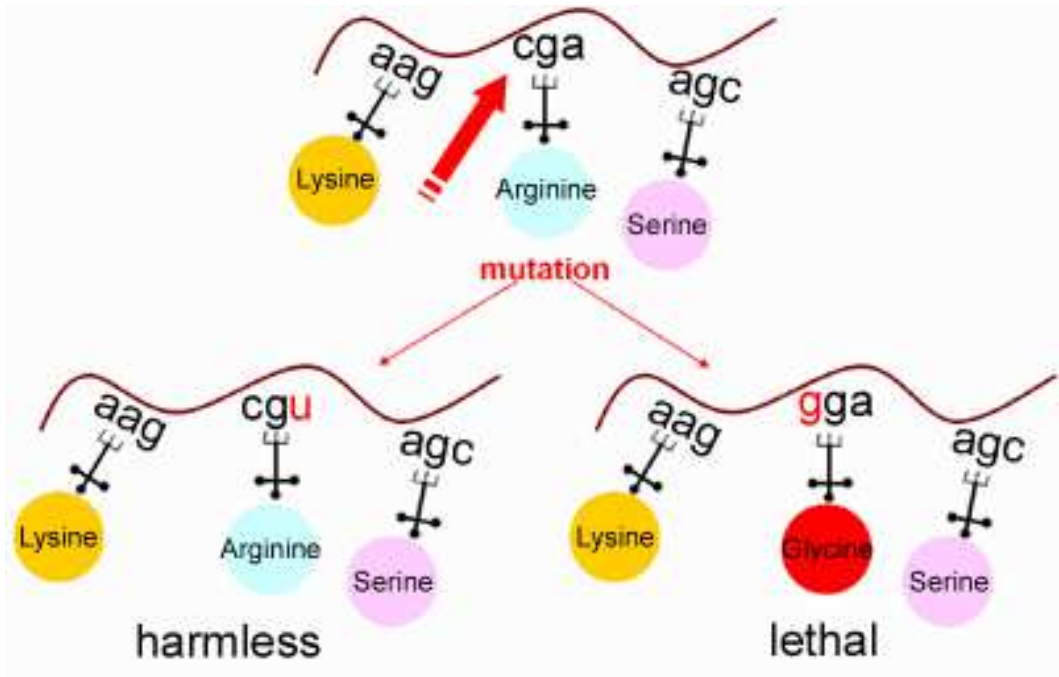}
\caption{\label{mutation}A mutation on exon part may be either harmless or lethal depends on whether it change the coded amino acid.}

\end{figure}

When a mutation takes place on the exon part, there are two possibilities: The changed codon either will code the same amino acid or it will code a different amino acid  [Fig. \ref{mutation}] since an amino acid can be coded by more than one codon. If the mutant codon still codes for the same amino acid, this mutation is taken to be harmless because at the end it will not affect the synthesized protein. However, if the mutant codon codes for a different amino acid, the protein cannot be produced and the individual would simply die.

To be more explicit, the codons AAA and AAG code the same amino acid, ``lysine''; hence if AAA turns into AAG as a result of a mutation the amino acid will not change and the protein can be constructed safely. However; if AAA turns into AGA, which codes the amino acid ``arginine'', the amino acid chain will change and we assume that the protein can not build up, which means the represented organism will die. There can be a mutation which converts AAA to AAX where X $\neq$ {A, G, C, or T}; then the individual dies automatically. As a model, we are looking at a simpler case where a mutation changes A to one of G, C, or T, but not X.

Since reproduction is not included in the model, the population can only diminish. The decrease in population can be found by calculating the probability of a deleterious mutation. The probability of the mutation changing the amino acid depends on  the codon; so one needs to find the probability of hitting each different codon type. First, the probability of hitting a codon type ($P_{\alpha}$) is calculated as the ratio of the number of codons of that type in the gene ($N_{\alpha}$) to total number of codons. Then we need to exclude the mutations that do not cause a change in the amino acid and calculate the probability of a change occurring in the amino acid caused by
a change in one nucleotide ($P(d/{\alpha}$)). 

We used only the exon (protein coding) part of the gene considering any mutations on the intron part is harmless. As a simple example, the human cytokine gene has a total length of 2068 nucleotides; 621 nucleotides in exon part and 1447 ones in intron. The probability of hitting the exon part of the gene is simply the ratio of the exon part to the total gene:
\begin{equation}
P(hitting \: exon) = \frac{621}{2068}= 0.3032\label{}
\end{equation}

\noindent Hence; the probability of having a \textit{deleterious} mutation for all of the gene is simply a product of mutation probability  and probability of hitting the exon part of gene. As the chances of hitting any part of the gene is a same, we can neglect the intron part in the simulation since this would only be a multiplicative constant in the problem. Therefore the probability of having a deleterious mutation for all of the human cytokine gene is simply:

\begin{equation}
P(deleterious) \propto \sum_{\alpha = 1}^{64} [
P_{\alpha} P(d/\alpha)]  = 0.7729 \label{}
\end{equation}

\noindent The survival probability can be calculated by:

\begin{equation}
P(surviving) = 1- P(deleterious) = 0.2271 \label{}
\end{equation}

If we take an initial population of $N_0$ genes (individuals), after $n$ number of mutations, to the first order, the number of surviving individuals ($N_n$) is given by:

\begin{equation}
N_n  \approx N_0P(surviving)^n \label{}
\end{equation}

\noindent Hence, we obtain the ``probability of survival" with the slope of the number of surviving individuals versus time graph:

\begin{equation}
slope \approx ln[P(surviving)] = -1.4823 \label{slope}
\end{equation}

Similarly the probability of survival can be calculated for all the genes separately. However in this calculation once we make a change in the gene sequence and if the individual survives, we forget about the change we have made and restart the process for the second mutation cycle with the original gene sequence. In Nature, if the individual survives, the second mutation cycle starts with the mutated gene sequence and not the original one. Therefore, to be able to get closer to Nature we have also written a simulation code which allows for the mutation in the gene sequence to be kept in the next mutation mutation cycle.

\section{SIMULATION\protect\\ }
\label{sec:level3}

In this simulation, the population consists of individuals which are described by only one gene. Genes are represented by arrays which contain 0, 1, 2, and 3's instead of the nucleotides Adenine (A), Guanine (G), Cytosine (C) and Thymine (Uracil (U)) respectively. A sign bit which shows if the gene has a deleterious mutation (1) or not (0) is also included in the array \cite{Gultepe05}.

In every time step, all of the individuals undergo a random mutation. If the mutation is deleterious, i. e. if it changes the amino acid, the sign bit is changed to `1', the individual is deleted from the population (death) and the time step is recorded. Otherwise, the sign bit is kept `0' and the individual survives.

When the mutation cycle is finished, the number of surviving individuals in each time step is calculated. Since the probability of mutation is independent of the number of individuals, surviving individuals also give us the population size. Hence, we have an exponential population decay and the exponent depends on the probability of surviving ($P(surviving)$). Logarithm of the population is fitted to a straight line and the slope of the line is calculated. All simulations are run for 10 times and probability of surviving is calculated according to the weighted average of these 10 runs.

\section{RESULTS AND DISCUSSION\protect\\ }
\label{sec:level4}

In this work, we have investigated the following amino acid tables in addition to \textit{SGC}:

\begin{itemize}
\item{Alternative Yeast Nuclear Code (AYNC)},
\item{Ascidian Mitochondrial Code (AMC)},
\item{Blepharisma Nuclear Code (BNC)},
\item{Ciliate, Dasycladacean and Hexamita Nuclear Code (CDHNC)},
\item{Echinoderm Mitochondrial Code (EMC)},
\item{Euplotid Nuclear Code (ENC)},
\item{Flatworm Mitochondrial Code (FMC)},
\item{Invertebrate Mitochondrial Code (IMC)},
\item{Mold, Protozoan, and Coelenterate Mitochondrial Code and Mycoplasma/Spiroplasma Code (MSC)},
\item{Vertebrate Mitochondrial Code (VMC)}, and
\item{Yeast Mitochondrial Code (YMC)}.
\end{itemize}

The results of our simulations, using two different human gene sequences, are given in Table \ref{results1} and Table \ref{results2}. We have selected these two genes as representatives, however, the same results have been obtained in many other human genes simulated. We have also used some genes from Mus Musculus (common house mouse) and Rattus (rat) where we have obtained similar results. As these detailed results are more appropriate for an evolutionary biology journal, we report only the representative results.

\begin{table}[!h]
\caption{\label{results1}Average slopes of the population decrease comparing different genetic code tables with human cytokine gene. Larger the magnitude of the slope is, the more the survival chance.} 

\begin{tabular}{|c|c|c|}

\hline Code & Simulation & Calculation\\
\hline FMC & $-1.4056 \pm 0.0005$ & 1.4053\\
\hline EMC & $-1.4164 \pm 0.0005$ & 1.4163\\
\hline IMC & $-1.4320 \pm 0.0001$ & 1.4319\\
\hline ENC & $-1.4415 \pm 0.0005$ & 1.4409\\
\hline BNC & $-1.4691 \pm 0.0003$ & 1.4685\\
\hline MSC & $-1.4759 \pm 0.0003$ & 1.4755\\
\hline CDHNC & $-1.4784 \pm 0.0005$ & 1.4779\\
\hline AMC & $-1.4784 \pm 0.0005$ & 1.4779\\
\hline \textbf{SGC} & $\textbf{-1.4830} \pm \textbf{0.0005}$ & $\textbf{1.4826}$\\
\hline VMC & $-1.5020 \pm 0.0001$ & 1.5017\\
\hline YMC & $-1.5677 \pm 0.0007$ & 1.5638\\
\hline AYNC & $-1.5800 \pm 0.0001$ & 1.5793\\
\hline
\end{tabular}
\end{table}

\begin{table}[!h]
\caption{\label{results2}Average slopes of the population decrease comparing different genetic code tables with human ARNT gene. Larger the magnitude of the slope is, the more the survival chance.} 

\begin{tabular}{|c|c|c|}

\hline Code & Simulation & Calculation\\
\hline FMC & $-1.4038 \pm 0.0004$ & 1.4036\\
\hline EMC & $-1.4101 \pm 0.0007$ & 1.4099\\
\hline CDHNC & $-1.4268 \pm 0.0004$ & 1.4268\\
\hline IMC & $-1.4288 \pm 0.0004$ & 1.4285\\
\hline BNC & $-1.4368 \pm 0.0004$ & 1.4368\\
\hline ENC & $-1.4566 \pm 0.0005$ & 1.4565\\
\hline AMC & $-1.4586 \pm 0.0005$ & 1.4583\\
\hline MSC & $-1.4609 \pm 0.0006$ & 1.4607\\
\hline \textbf{SGC} & $\textbf{-1.4653} \pm \textbf{0.0005}$ & $\textbf{1.4650}$\\
\hline VMC & $-1.4882 \pm 0.0008$ & 1.4878\\
\hline AYNC & $-1.5215 \pm 0.0005$ & 1.5213\\
\hline YMC & $-1.5295 \pm 0.0005$ & 1.5271\\
\hline
\end{tabular}
\end{table}

If it is assumed that the genetic code table is optimized for increasing chance of survival against mutations, then the results of Table \ref{results1} and Table \ref{results2} cause many concerns. Even though there are few code tables giving results which favor the usage of \textit{SGC} (like VMC, AYNC, and YMC), we can see that if we use a different code table, for example \textit{FMC}, our white blood cell production would be more resilient towards mutations.

To be certain that the results obtained in Table \ref{results1} and Table \ref{results2} do not depend on particular genes, we have first created an artificial average human gene and we run simulations using this average human gene as representetive of individuals. Table \ref{results3} shows the results from these simulations.

\begin{table}[h]
\caption{\label{results3}Average slopes of the population decrease  using different genetic code tables and the average  human gene. Larger the magnitude of the slope is, the more the survival chance.} 
\begin{tabular}{|c|c|c|}

\hline Code & Simulation & Calculation\\
\hline FMC & $-1.4081 \pm 0.0005$ & 1.4083\\
\hline EMC & $-1.4204 \pm 0.0005$ & 1.4207\\
\hline IMC & $-1.4438 \pm 0.0003$ & 1.4439\\
\hline CDHNC & $-1.4498 \pm 0.0001$ & 1.4501\\
\hline BNC & $-1.4554 \pm 0.0006$ & 1.4558\\
\hline ENC & $-1.4595 \pm 0.0005$ & 1.4596\\
\hline MSC & $-1.4655 \pm 0.0005$ & 1.4658\\
\hline AMC & $-1.4693 \pm 0.0006$ & 1.4697\\
\hline \textbf{SGC} & $\textbf{-1.4712} \pm \textbf{0.0004}$ & $\textbf{1.4716}$\\
\hline VMC & $-1.4969 \pm 0.0005$ & 1.4971\\
\hline YMC & $-1.5540 \pm 0.0006$ & 1.5543\\
\hline AYNC & $-1.5803 \pm 0.0001$ & 1.5804\\
\hline
\end{tabular}
\end{table}

In Table \ref{results3}, \textit{FMC} still performs much better and the worst performance is still by \textit{AYNC} followed by \textit{YMC}. The human average gene gives comparable results to the genes chosen in this work, indicating that the advantage of \textit{SGC} cannot be simply explained on the basis of singular mutations without taking higher order effects like protein folding into account.

\section{CONCLUSION\protect\\ }
\label{sec:level6}

In this paper, we used a computer simulation to represent a living organism under mutations. Furthermore, we changed the genetic code used in the simulations to analyze its effect on population stability.

We have used different code tables utilized in the Nature in connection with two sample genes, human cytokine gene and human ARNT gene . Since these genes are one of the key factors for our bodies, we assumed that if the organism fails to produce any of the proteins coded by these genes, it will not be able to survive.

The simulations show that \textit{SGC}, which is being used in most of the vertebrates, actually does not give the most resilient organisms against point mutations if we only apply simple simulation rules.

To test these results, we have also created an artificial average human gene. If we compare our results as \textit{SGC} versus the rest of the coding schemes, the results do not change, i. e., \textit{SGC} is still not the best solution. However, when we compare other code tables within themselves we see that some code tables give better results with different human genes whereas others give better results with the artificial average human gene. This result is most pronounced with \textit{CDHNC}, which survives much better if we use the average human gene than the human cytokine gene.

\section{ACKNOWLEDGEMENTS}
I am grateful to Dr. Isil Aksan Kurnaz and Dr. Muhittin Mungan for their contributions on the model and the calculations. This work has been supported by Bogazici University under BAP 07B302.


\begin{thebibliography}{10}
\expandafter\ifx\csname bibnamefont\endcsname\relax
  \def\bibnamefont#1{#1}\fi
\expandafter\ifx\csname bibfnamefont\endcsname\relax
  \def\bibfnamefont#1{#1}\fi
\expandafter\ifx\csname url\endcsname\relax
  \def\url#1{\texttt{#1}}\fi
\expandafter\ifx\csname urlprefix\endcsname\relax\def\urlprefix{URL }\fi
\providecommand{\bibinfo}[2]{#2}
\providecommand{\eprint}[2][]{\url{#2}}

\bibitem{Heumann1995}
\bibinfo{author}{\bibfnamefont{N. G. F.}  \bibnamefont{Medeiros}} \bibnamefont{and}
  \bibinfo{author}{\bibfnamefont{R. N.} \bibnamefont{Onody}}
  \bibinfo{journal}{Phys. Rev. E} \textbf{\bibinfo{volume}{64}},
  \bibinfo{pages}{041915} (\bibinfo{year}{2001}).

\bibitem{Mueller1996}
\bibinfo{author}{\bibfnamefont{L. D.}  \bibnamefont{Mueller}} \bibnamefont{and}
  \bibinfo{author}{\bibfnamefont{M. R.} \bibnamefont{Rose}}
  \bibinfo{journal}{PNAS} \textbf{\bibinfo{volume}{93}},
  \bibinfo{pages}{15249} (\bibinfo{year}{1996}).

\bibitem{Cui2000}
\bibinfo{author}{\bibfnamefont{Y.}  \bibnamefont{Cui}}
  \bibinfo{author}{\bibfnamefont{R. S.} \bibnamefont{Chen}} \bibnamefont{and}
 \bibinfo{author}{\bibfnamefont{W. H.}  \bibnamefont{Wong}}
  \bibinfo{journal}{PNAS} \textbf{\bibinfo{volume}{97}},
  \bibinfo{pages}{3330} (\bibinfo{year}{2000}).

\bibitem{Medeiros2001}
\bibinfo{author}{\bibfnamefont{M.}  \bibnamefont{Heumann}} \bibnamefont{and}
  \bibinfo{author}{\bibfnamefont{M.} \bibnamefont{Hötzel}}
  \bibinfo{journal}{J. Stat. Phys.} \textbf{\bibinfo{volume}{79}},
  \bibinfo{pages}{483} (\bibinfo{year}{1995}).


\bibitem{Penna}
\bibinfo{author}{\bibfnamefont{T. J. P.} \bibnamefont{Penna}},
  \bibinfo{journal}{J. Stat. Phys.} \textbf{\bibinfo{volume}{78}},
  \bibinfo{pages}{1629} (\bibinfo{year}{1995}).

\bibitem{Weaver2002}
\bibinfo{author}{\bibfnamefont{R. F.}~\bibnamefont{Weaver}},
  \emph{\bibinfo{title}{Molecular Biology}} (\bibinfo{publisher}{McGraw-Hill, New York}, \bibinfo{year}{2002}).

\bibitem{NCBIcodes}
\url{www.ncbi.nlm.nih.gov/Taxonomy/Utils/wprintgc.cgi?mode=c}.

\bibitem{Gultepe05}
\bibinfo{author}{\bibfnamefont{E.} \bibnamefont{Gultepe}} \bibnamefont{and}
  \bibinfo{author}{\bibfnamefont{M. L.} \bibnamefont{Kurnaz}}
  \bibinfo{journal}{Physica A} \textbf{\bibinfo{volume}{357}},
  \bibinfo{pages}{525} (\bibinfo{year}{2005}).

\bibitem{gene}
\url{www.ncbi.nlm.nih.gov/entrez/viewer.fcgi?db=nucleotide\&val=285912}.


\bibitem{Kazusa}
\url{www.kazusa.or.jp/codon/cgi-bin/showcodon.cgi?species=Homo+sapiens +[gbpri]}

\bibitem{Jukes21920}
\bibinfo{author}{\bibfnamefont{T. H.} \bibnamefont{Jukes}} \bibnamefont{and}
  \bibinfo{author}{\bibfnamefont{C. R.} \bibnamefont{Cantor}}
  \emph{\bibinfo{title}{in Mammalian Protein Metabolism, edited by H. N. Munro}} 
  (\bibinfo{publisher}{Academic Press, New York}, 
  \bibinfo{pages}{21} \bibinfo{year}{1969}).

\end{thebibliography}
\end{document}